\documentstyle[preprint,aps,prl]{revtex}

\newcommand{\be}{\begin{equation}}
\newcommand{\ee}{\end{equation}}
\newcommand{\ba}{\begin{eqnarray}}
\newcommand{\ea}{\end{eqnarray}}
\newcommand{\ket}[1]{| {#1} \rangle}

\def\lsim{\mathrel{\rlap{\lower4pt\hbox{\hskip1pt$\sim$}}
    \raise1pt\hbox{$<$}}}     
\def\gsim{\mathrel{\rlap{\lower4pt\hbox{\hskip1pt$\sim$}}
    \raise1pt\hbox{$>$}}}     

\begin{document}

\draft

\title{Spectral Decorrelation of Nuclear Levels
in the Presence of Continuum Decay}

\author{S. Dro\.zd\.z\,}

\address{ Institute of Nuclear Physics, PL - 31-342 Krak\'ow, Poland\\
Institut f\"ur Kernphysik, Forschugszentrum J\"ulich,
D-51425 J\"ulich, Germany}

\author{A. Trellakis\cite{byline2} and J. Wambach\cite{byline2}}

\address{Department of Physics, University of Illinois at Urbana,
Illinois 61801}

\maketitle

\begin{abstract}
The fluctuation properties of nuclear giant resonance spectra are studied
in the presence of continuum decay.
The subspace of quasi-bound states is specified by one-particle
one-hole and two-particle two-hole excitations and the continuum coupling
is generated by a scattering ensemble.
It is found that, with increasing number of open channels, the real parts of
the complex eigenvalues quickly decorrelate. This appears to be related
to the transition from power-law to exponential time behavior of the survival
probability of an initially non-stationary state.
\end{abstract}


Level fluctuations, measured in terms of the
nearest-neighbor-spacing-distribution (NNSD)
and the $\Delta_3$-statistics, provide a commonly accepted tool for
studying the quantum interplay between regular and chaotic dynamics. The
standard treatment is restricted to bound states while,
in many cases, the excited states are resonances embedded in the
continuum. Already a generalization of the standard two-level repulsion
theorem \cite{NW} to resonances \cite{Bre} shows that this may significantly
modify the correlations between the states.
Generically, chaotic dynamics leads to level repulsion but the presence
of the continuum (open system), is expected \cite{MZ} to wash out the
repulsion between the resonance energies.
On the other hand, the lack of correlations between levels is normally
interpreted as a manifestation of regular dynamics. It thus seems necessary to
explore, on a fully quantitative level, what is the nature of the weakening of
the repulsion due to openness and how it modifies the fluctuation
characteristics.

The most practical way for describing an irreversible decay into the continuum
is based on a scattering ensemble of non-hermitian random matrices \cite{SZ}.
Such a treatment follows naturally from the
projection-operator technique \cite{Fes} in which the subspace of
asymptotically decaying states is formally eliminated. The resulting
non-hermitian Hamiltonian
\begin{equation}
{\cal H}=H - {i\over 2} W
\label{eq:hnonh}
\end{equation}
acts in the space of quasi-bound states and the coupling to the
continuum is accounted for by the anti-hermitian operator $W$. Unitarity of
the scattering matrix imposes on $W$ the following factorization condition:
\begin{equation}
W={\bf A}{\bf A}^T.
\label{eq:W}
\end{equation}
For an open quantum system with $N$ quasi-bound states, $\ket{i}$,
($i=1,...,N$) which decay into $k$ open channels $a$
($a=1,...,k$), the $N \times k$ matrix ${\bf A}\equiv\{A^a_i\}$ denotes the
amplitudes for connecting the states $|i\rangle$ to the reaction channels $a$.
The diagonalization of $\cal H$ in the basis $\ket{i}$ yields $N$
quasi-stationary states with complex eigenvalues
${\cal E}_j = E_j - i \Gamma_j/2$, whose imaginary parts correspond to the
'escape width'. The factorization of $W$ guarantees that $\Gamma_j \ge 0$.
An interesting effect \cite{SZ,Rot} -- due to the separable form of $W$ --
is that, in the strong-coupling limit ($W\gg H$), one observes a segregation
of the states: $k$ states
accumulate most the total width, $\Gamma = \sum_j \Gamma_j$, while the
remaining $N-k$ states have nearly vanishing widths (they become 'enslaved'
\cite{Rot}).

For systems, such as the atomic nucleus, whose dynamics is expected
to be classically chaotic, it is natural to consider the
hermitian- and the anti-hermitian parts of $\cal H$ to be
statistically independent \cite{SZ}. Furthermore, the real and symmetric
$N \times N$ matrix $H$
can be modeled \cite{SZ} as a member of the Gaussian orthogonal ensemble
(GOE) of random matrices \cite{BFF,Boh}. For large $N$ the matrix elements
of $H$ obey the following pair contraction formula:
\begin{equation}
\langle H_{ii'}H_{jj'} \rangle ={a^2 \over 4N}
(\delta_{ij}\delta_{i'j'} + \delta_{ij'}\delta_{i'j})
\label{eq:pair}
\end{equation}
in the sense of GOE averaging. The constant $a$ is related to the mean level
spacing, $D=2a/N$.

For a general Gaussian ensemble of complex random matrices $\cal H$ \cite{Gin}
an analogous contraction formula for
$\langle {\cal H}_{ij} {\cal H}_{i'j'} \rangle$ is obtained which implies that
the real and imaginary parts of $\cal H$ commute on average.
Consequently, the two hypersurfaces,
representing the real and imaginary parts of the energy lie in orthogonal
subspaces \cite{MH}. This, for sufficiently large $N$, may produce decorrelated
spectra as seen from either the real or imaginary axes, in spite of a cubic
repulsion on the complex plane.

However, this general Gaussian ensemble of complex random matrices
is not applicable in the present case because of $S$-matrix unitarity. Instead,
the anti-hermitian part of $\cal H$ is determined by the
amplitudes $A^a_i$ via Eq.~(\ref{eq:W}). Based on the GOE character of internal
dynamics and orthogonal invariance arguments \cite{SZ} the amplitudes
$A^a_i$ can be assumed to be Gaussian distributed.
The corresponding correlator reads:
\begin{equation}
\langle A^a_i A^b_j \rangle = {1\over N} \gamma^a \delta^{ab}
\delta_{ij},~~~~~~~~~~ \langle A^a_i \rangle = 0
\label{eq:Acor}
\end{equation}
implying that the average trace is
$\langle Tr W \rangle =
\Sigma_a \gamma^a$. The diagonal elements $W_{ii}=\Sigma_{a=1}^k
(A^a_i)^2$ are then positive, statistically independent and
obey a $\chi_k$-square distribution.

Unlike the amplitudes $A^a_i$ the matrix elements of $W$ are not
statistically independent, however. The number of
independent random parameters, $Nk - {1\over 2} k(k-1)$ for $k \le N$, is
reduced by the second term as a consequence of the rotational invariance of
$W_{ij}=\Sigma_{a=1}^k A^a_i A^a_j$ (the scalar product
between $N$ $k$-dimensional vectors ${\bf A}_i$ in the channel space).
Only for $k=N$ the correlations in $W$ are specified by
${1\over 2} N(N-1)$ parameters, as for the GOE. Thus a
decorrelation of the projected spectra may result.
In most realistic cases, however, the number of open channels $k$ is smaller
than $N$. To assess the dependence on the number of open channels we perform
a systematic numerical study of the spectral correlations as a function of $k$.

Since the nuclear interaction is predominantly two body in nature,
the matrix representation of the nuclear Hamiltonian should be related to the
so-called 'embedded' Gaussian orthogonal ensemble (EGOE) \cite{BFF} rather than
the GOE. Therefore, to make our study realistic from the nuclear physics point
of view, we generate the hermitian part of $\cal H$ from the model in
ref.~\cite{DNSW} instead of using a GOE random ensemble. The
Hamiltonian includes a mean-field part and a zero-range and density-dependent
two-body interaction. The matrix representation of $H$ is expressed in the
basis of one-particle one-hole (1p1h) and two-particle two-hole (2p2h)
excitations generated by the mean-field part and by discretizing the continuum
\cite{DNSW}.
The spectral fluctuations of the corresponding real
eigenvalues, measured in terms of the NNSD and $\Delta_3$, coincide
with those of the GOE \cite{DNSW}, even though significant deviations from the
Gaussian distribution of the matrix elements are found \cite{TDW,Fla}.

Because of time-reversal invariance the anti-hermitian part of $\cal H$ is
generated by a Gaussian ensemble of real
amplitudes $A^a_i$ with correlator (\ref{eq:Acor}), where $\gamma^a=1$,
{\it i.e.} we assume that all channels are equivalent and the strength of the
external coupling is comparable to the internal one.
In the specific calculations presented below, we select quadrupole excitations
($J^{\pi}=2^+$) in $^{40}$Ca. To ensure acceptable statistics,
in the quasi-bound-state space all 1p1h and 2p2h states up to an excitation
energy
of 40 MeV are included. This yields a $1661\times 1661$ Hamiltonian matrix.
Fig.~1 shows the resulting eigenvalue distribution on the complex energy plane
for an increasing number $k$ of open channels. For $k=10$ the
majority of the energies lie very close to the real axis and only a few
states acquire a significant width which is a trace of the
'collective synchronization' discussed in ref.~\cite{SZ,Rot}.
Increasing $k$, the distribution becomes more uniform and
the width $\Delta_g$ of the empty strip between the cloud of eigenvalues and
the real axis widens. This is understandable as $\Delta_g$ is equal
to the 'correlation width' which describes the asymptotic
behavior of the decay process \cite{LSSS}.

The NNSD on the plane can be determined by calculating the normalized
distances $s_i=d_i \rho_n({\cal E}_i)^{1/2}$, where $d_i$ stands for the
Euclidean
distance  between the eigenvalue ${\cal E}_i$ and its nearest
neighbor, and $\rho_n({\cal E}_i)$ for the local density of eigenvalues
determined from $n$ nearest neighbors of ${\cal E}_i$.
Similarly as in ref.~\cite{HIL},
the choice $n=10$ turns out satisfactory and guarantees stability.
The numerical results are compared to the Poisson distribution
$P(s)=(\pi/2) s \exp(-\pi s^2/4)$ (dashed lines in the {\sl rh} column of
Fig.~1), which shows linear repulsion on the plane, and to the
$P(s)= (81\pi^2/128) s^3 \exp(-9\pi s^2/16)$ with cubic
repulsion (solid lines).
The latter gives a good description for the NNSD of symmetric Gaussian
random matrices \cite{TDW}\cite{JMSS} and, for a large number of
open channels, also  fit our numerical results nicely. For a few open channels
(upper right part of Fig~1.) we see a weaker then cubic repulsion, however .

Now we come to the central point namely the fluctuation properties
of the real parts $E_i$ of the energy eigenvalues. The corresponding NNSD and
$\Delta_3$-statistics are shown in Fig.~2. It is well known that, without
coupling the continuum,  the spectra show GOE characteristics for both
measures \cite{DNSW}. However, for many open channels a decorrelation takes
place. In fact, for large $k$ the results are well reproduced by a
Poissonian shape of the NNSD (lower left part of Fig.~2). Quite surprisingly,
this even holds for $k/N$ of a few percent (middle left part of Fig.~2).
Already for ten open channels $(k/N=6*10^{-1})$, there is a visible deviation
from the Wigner distribution (upper left part of Fig.~2). These numerical
observations lead to the conclusion that the appropriate way of describing
these deviations is to consider superpositions of Wigner and Poisson
distributions rather than Wigner and Gaussian \cite{MZ}.

The longer-range correlations (spectral rigidity) expressed by the
$\Delta_3$-statistics show a similar tendency, although the transition is
somewhat slower. In addition, as is seen in Fig.~2, the transition region
$L_{max}$ from GOE
to Poissonian characteristics is restricted to about 10 normalized
distance units. This appears to be consistent with the findings in \cite{DS}
for hermitian separable problems, where $L_{max}$ increases
with increasing length of the string of eigenvalues. In the present
case the string is comparatively short. On a more formal level \cite{Berry},
the $\Delta_3$-statistics is known to be non-universal above a certain
$L_{max}$. For systems with a known classical limit, $L_{max}$ is determined
by the inverse of the period of the shortest periodic orbits. We wish to
mention, without showing the results explicitly, that an analogous analysis
for the imaginary parts of ${\cal E}_i$ show Poissonian fluctuations for any
number of the open channels.
This asymmetry in the statistical properties of $E_j$ and $\Gamma_j$
is related to the different properties of the real and imaginary parts of
$\cal H$, especially for smaller values of $k$.

Another way of understanding the decorrelation of the resonance energies due
to the presence of continuum decay comes from the relation
between the wave-packet
dynamics and the stationary states \cite{Heller}. The
latter can be obtained via the Fourier transform of the time evolution of a
generic wave packet. For a bound-state problem such a wave packet resides in
the interaction region forever and thus, the structure of the corresponding
phase space can be resolved with arbitrary accuracy. Consequently, for a
chaotic system, the whole complexity (delocalization, random nodal pattern,
scars, etc.) of stationary states can be reproduced. Coupling to the continuum,
sets a limit for this process, however. As time progresses, the wave packet
will leak out of the interaction region and makes it impossible
to resolve all details of the dynamics. As a result the wave
functions, projected onto the interaction region, look more regular than their
counterparts in a closed system. The leakage is expected to occur faster with
increasing $k$. A quantititive measure of the speed is the survival
probability $P(t)$ of a randomly chosen wave packet $\ket{F}$, initially
localized in the interaction region. As a convenient and experimentally
motivated choice we consider a state excited by the
isovector quadrupole operator $(|F\rangle={\hat F} |0\rangle$).
When expanded $\ket{F}$ involves all the eigenstates $|\chi_i\rangle$
of $\cal H$  and
\begin{equation}
P(t)=|\langle F(0)|F(t)\rangle|^2= |\sum_{j=1}^N \langle 0|\hat F|\chi_j\rangle
\langle \chi_j|\hat F|0\rangle e^{i {\cal E}_j t/\hbar}|^2
\label{eq:P}
\end{equation}
(for a complex symmetric matrix the left and right eigenvectors
are the same).
In the absence of continuum coupling, $P(t)$ remains
constant (on average) after a rapid initial dephasing
due to the non-stationarity of $|F\rangle$ \cite{DNWS}.
For an open system, on the other hand, a decay of $P(t)$ is to be expected.
The most interesting feature is the dependence of the decay law on the
number of open channels: For a small $k$ the decay is very slow and well
represented by a power-law $(P(t) \sim t^{-z})$. For $k=1$ we find
$z\approx -1/2$, in reasonable agreement with the estimates of
ref.~\cite{DHM}.
As $k$ increases $z$ grows very fast and, for $k>100$, $P(t)$ drops
exponentially on long time scales, {\sl i.e.}  $P(t) \sim \exp(-\eta t)$,
with the decay constant $\eta$ growing rapidly with $k$ (Fig.~3).
These observations go in parallel with the classical picture of open
phase space phenomena such as a chaotic scattering \cite{DOS}: For a small
number of the open channels the decay is governed by a power-law. This is
associated with larger fractal dimensions of the set of singularities
generating chaotic
behavior than for many open channel cases which lead to an exponential decay.

In summary, the numerical analysis presented in this work shows that
GOE correlated spectra of quasi-bound states become fully decorrelated in the
presence of continuum coupling and when the number of open channels is large.
This transition is accompanied  by a change of the decay properties of the
average survival probability of a non-stationary wave packet, turning from
power-law to exponential. This appears to be
consistent with the semiclassical relation \cite{BS}
between the time-dependence of $P(t)$ and the structure of the resonances.
An exponential behavior of $P(t)$ corresponds to the region of strongly
overlapping resonances (Ericson fluctuations \cite{Eri}),  while the
power-law decay, with small power indices $z$ \cite{LFO}, corresponds to
isolated resonances, and it is this isolation which preserves the original
fluctuations.
\vspace{0.5cm}

This work was supported in part by the Polish KBN Grant
No. 2 P302 157 04 and by a grant from the National Science Foundation,
NSF-PHY-94-21309.

\parindent=.0cm            


\newpage
\begin{center}
{\large \sl \bf Figure Captions}
\end{center}
\vspace{0.5cm}

\begin{itemize}

\item[{\bf Figure 1}:] Left column: The eigenvalue distribution of the
non-hermitian
Hamiltonian $\cal H$ defined in Eq.~(1) for different number $k$
of open channels. The hermitian part $H$ is chosen as the Hamiltonian of
[12] while the anti-hermitian part $W$ is given by  Eq.~(2)
taking the amplitudes $A$ as members of the Gaussian ensemble [4].
Right column: the corresponding NNSD on the complex plane.

\item[{\bf Figure 2:}]
The NNSD ({\sl lhs}) and the $\Delta_3$ statistics ({\sl rhs}) of the
real parts $E_i$ for energy eigenvalues of ${\cal H}$ and different number k
of open channels.

\item[{\bf Figure 3:}]
The time dependence of the survival probability $P(t)$ of a wave
packet, initialized by the isovector quadrupole operator, for various numbers
of open channels.

\end{itemize}

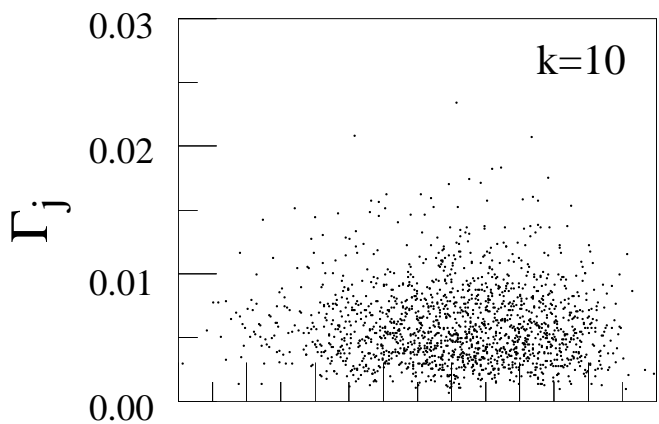
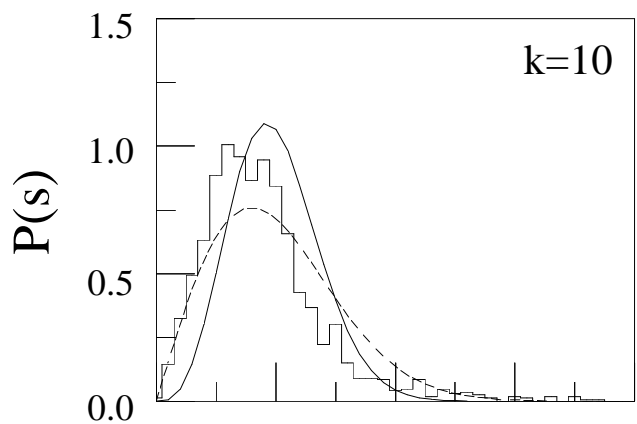
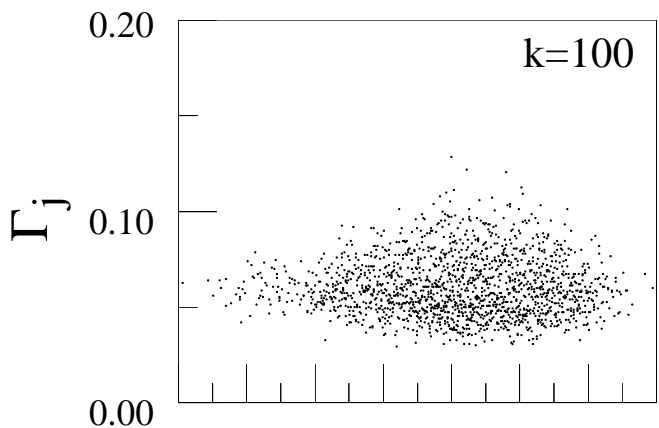
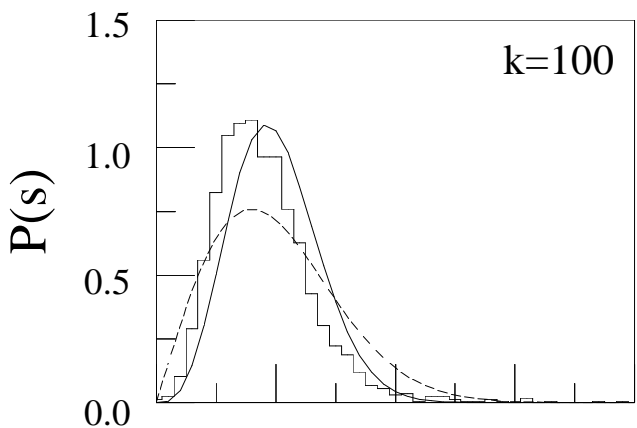
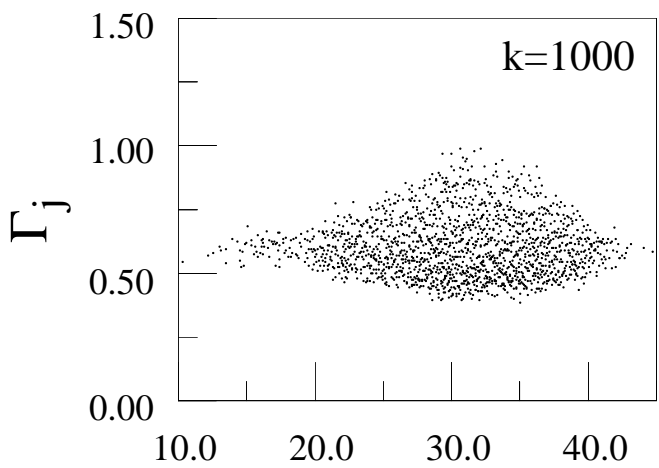
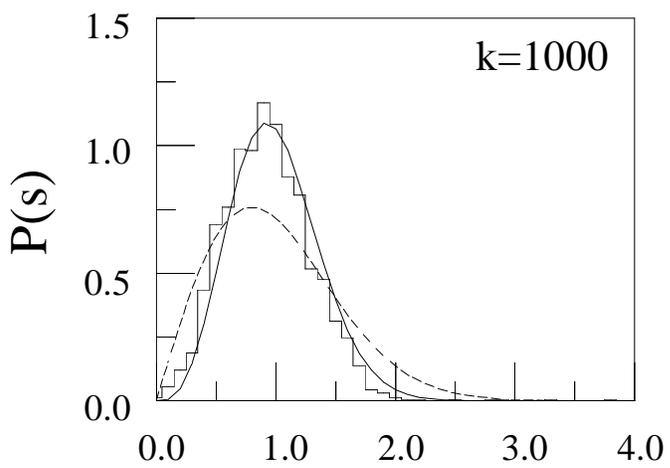
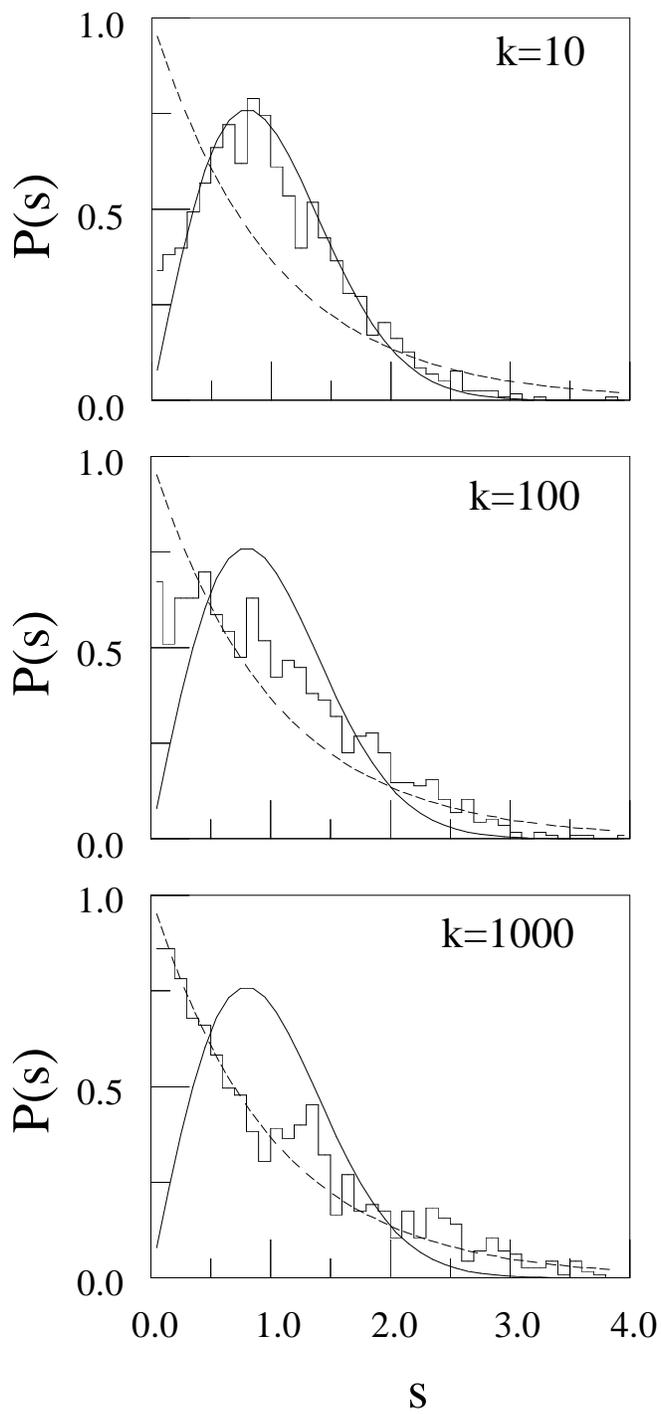
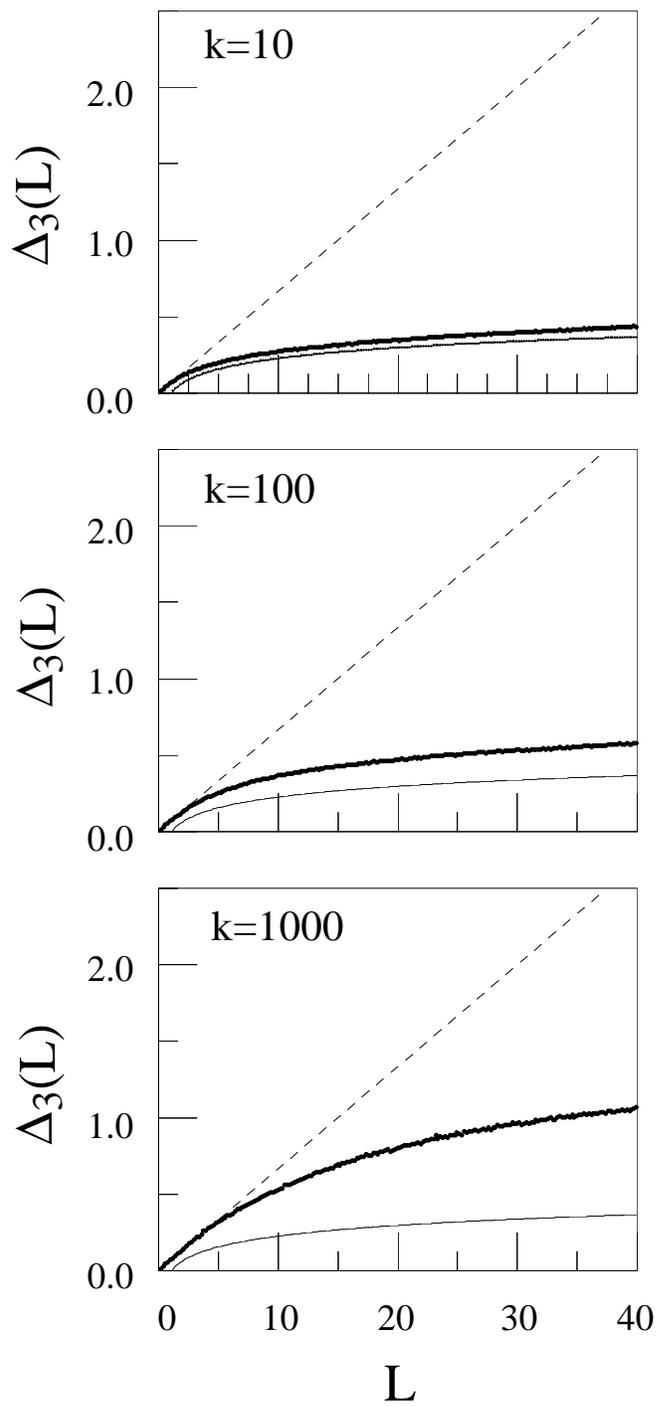
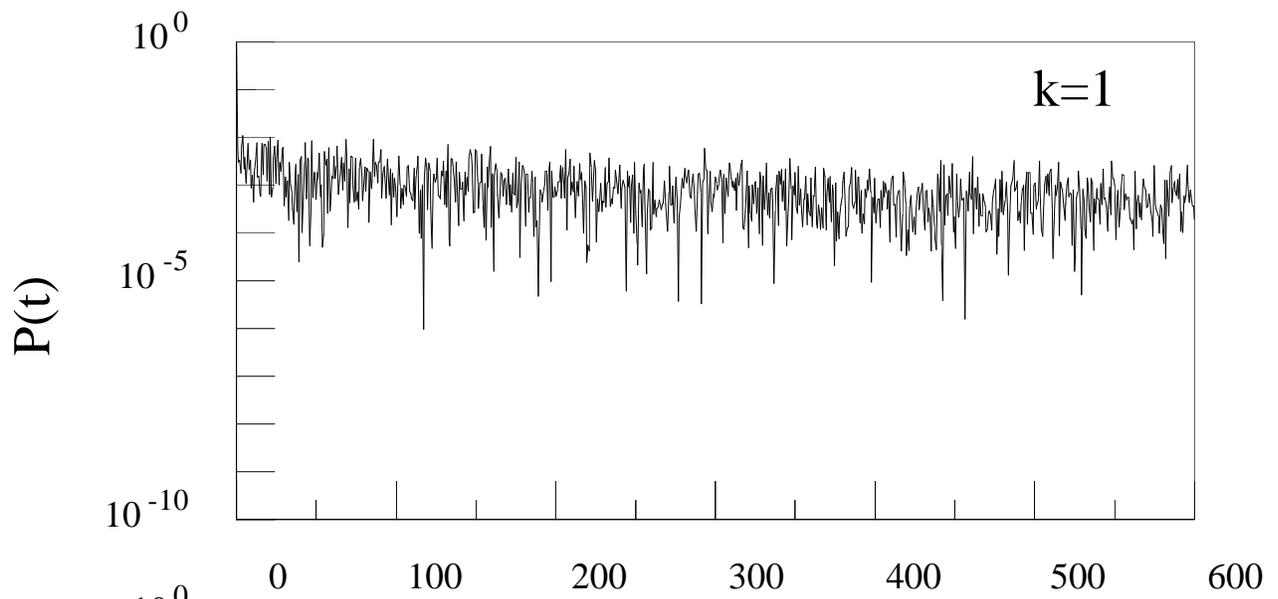
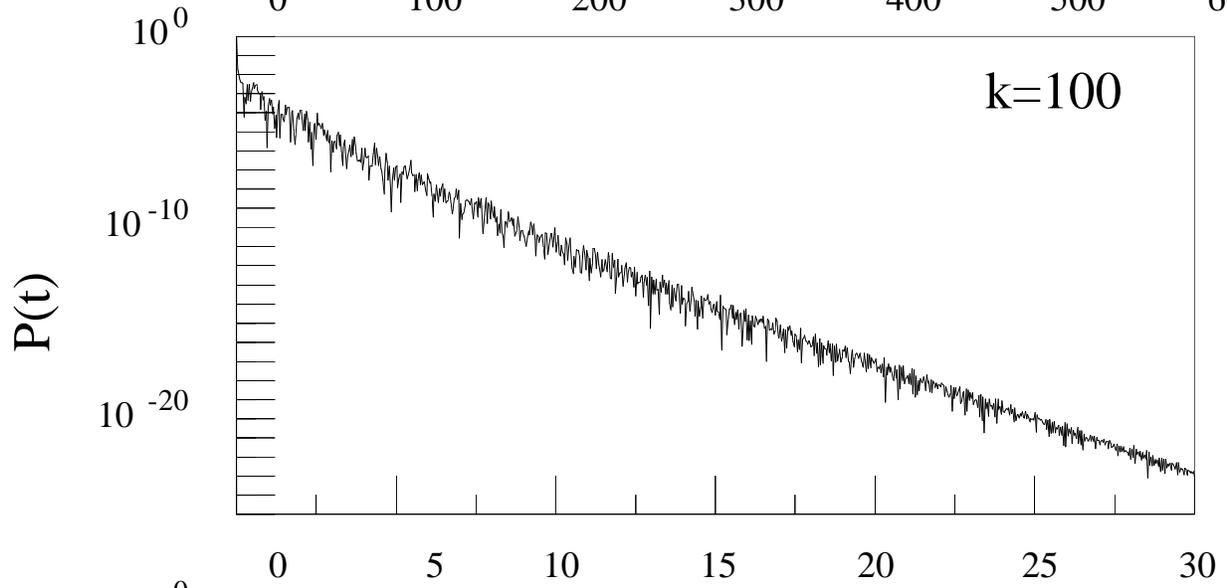
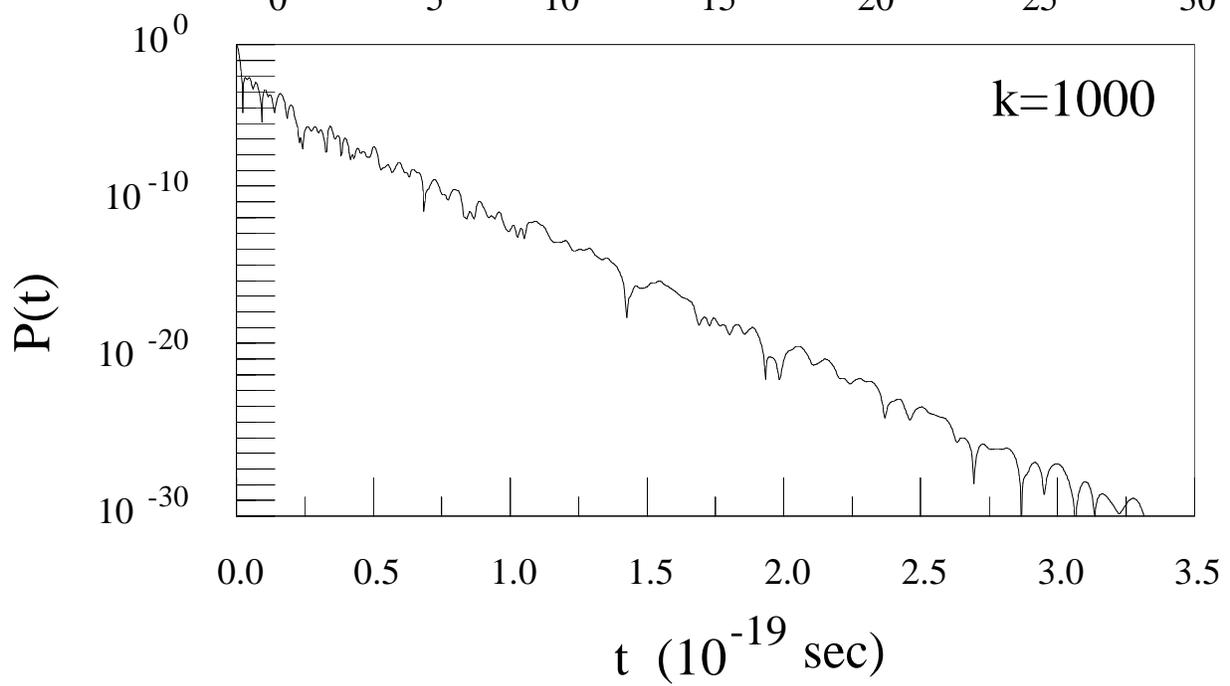
\end{document}